\documentclass{emulateapj}

 \newcommand{\lsim}{{\, \lower2truept\hbox{
${< \atop\hbox{\raise4truept\hbox{$\sim$}}}$}\,}}
\newcommand{\gsim}{{\, \lower2truept\hbox{
${> \atop\hbox{\raise4truept\hbox{$\sim$}}}$}\,}}

\shorttitle{SN\,2005nc and GRB\,050525A}
\shortauthors{M. Della Valle et al.}

\begin{document}
 
\title{Hypernova Signatures in the Late Rebrightening of GRB\,050525A}

\author{
  M. Della Valle\altaffilmark{1,2},
  D. Malesani\altaffilmark{3},
  J.S. Bloom\altaffilmark{4},
  S. Benetti\altaffilmark{5},
  G. Chincarini\altaffilmark{6,7},
  P. D'Avanzo\altaffilmark{6,8},
  R.J. Foley\altaffilmark{4},
  S. Covino\altaffilmark{6},
  A. Melandri\altaffilmark{9,10}, 
  S. Piranomonte\altaffilmark{10},
  G. Tagliaferri\altaffilmark{6},
  L. Stella\altaffilmark{10},
  R. Gilmozzi\altaffilmark{11},
  L.A. Antonelli\altaffilmark{10},
  S. Campana\altaffilmark{6},
  H.-W. Chen\altaffilmark{12},
  P. Filliatre\altaffilmark{13,14},
  F. Fiore\altaffilmark{9},
  D. Fugazza\altaffilmark{6},
  N. Gehrels\altaffilmark{15},
  K. Hurley\altaffilmark{16},
  I.F. Mirabel\altaffilmark{17},
  L.J. Pellizza\altaffilmark{18},
  L. Piro\altaffilmark{19},
  J.X. Prochaska\altaffilmark{20}
}

\altaffiltext{1}{INAF/Arcetri, largo E. Fermi 5, I-50125 Firenze, Italy}
\altaffiltext{2}{Kavli Institute for Theoretical Physics, UCSB, Santa
  Barbara, CA 93106-4030, USA}
\altaffiltext{3}{SISSA/ISAS, via Beirut 2-4, I-34014 Trieste, Italy}
\altaffiltext{4}{Department of Astronomy, 601 Campbell Hall, University
  of California, Berkeley, CA 94720, USA}
\altaffiltext{5}{INAF/Padova, vicolo Osservatorio 5, I-35122 Padova, Italy}
\altaffiltext{6}{INAF/Brera, via E. Bianchi 46, I-23807 Merate (Lc), Italy}
\altaffiltext{7}{Universit\`a degli studi di Milano-Bicocca,
  Dipartimento di Fisica, piazza delle Scienze 3, I-20126 Milano}
\altaffiltext{8}{Dipartimento di Fisica e Matematica, Universit\`a
  dell'Insubria, via Valleggio 11, I-22100 Como, Italy}
\altaffiltext{9}{Astrophysics Research Institute, John Moores
  University, 12 Quays House, Egerton Wharf, Birkenhead CH41, UK}
\altaffiltext{10}{INAF/Roma, via di Frascati 33, I-00040 Monteporzio
  Catone (Roma), Italy}
\altaffiltext{11}{European Southern Observatory,
  Karl-Schwarzschild-Str. 2 D-85748 Garching bei Munchen, Germany}
\altaffiltext{12}{Kavli Institute for Astrophysics and Space Research,
  MIT, Cambridge, MA 02139-4307, USA}
\altaffiltext{13}{Laboratoire Astroparticule et Cosmologie, UMR 7164, 11
  Place Marcelin Berthelot, F-75231 Paris Cedex 05, France}
\altaffiltext{14}{Service d'Astrophysique, CEA/DSM/DAPNIA, CE-Saclay,
  F-91911 Gif-sur-Yvette Cedex, France}
\altaffiltext{15}{NASA, Goddard Space Flight Center, Greenbelt, MD
  20771, USA}
\altaffiltext{16}{Space Sciences Laboratory, University of California,
  Berkeley, CA 94720-7450, USA}
\altaffiltext{17}{European Southern Observatory, Alonso de C\'ordova
  3107, Vitacura, Casilla 19001, Santiago 19, Chile}
\altaffiltext{18}{AIM (UMR 7158 CEA/CNRS/Universit\'e Paris 7), Service
  d'Astrophysique, CEA Saclay, F-91191 Gif-sur-Yvette Cedex, France}
\altaffiltext{19}{INAF/IASF sezione di Roma, via del Fosso del Cavaliere
  100, I-00113 Roma, Italy}
\altaffiltext{20}{Lick Observatory, 373 Interdisciplinary Sciences,
  University of California, Santa Cruz, CA 95064, USA}

\begin{abstract}
We report observations of GRB\,050525A, for which a Gemini North
spectrum shows its redshift to be $z = 0.606$. This is the third closest
long GRB discovered by \textit{Swift}. We observed its afterglow using
the VLT, Gemini and TNG telescopes to search for an associated SN. We
find that the early-time light curve is described by a broken power law
with a break at $t \sim 0.3$~d after the burst. About 5~d after the
burst, a flattening is apparent, followed by a further dimming. Both the
magnitude and the shape of the light curve suggest that a supernova was
emerging during the late decay of the afterglow. This supernova, dubbed
SN\,2005nc, had a rise time faster than SN\,1998bw and a long-lasting
maximum. A spectrum obtained about 20~d (rest-frame) after the GRB
resembles the spectrum of SN\,1998bw obtained close to maximum light.
\end{abstract}

\keywords{gamma rays: bursts --- supernovae: individual (GRB 050525A)}

\section{Introduction}

In nearly one decade of optical and infrared studies of gamma-ray
bursts (GRBs), it has been established that at least a large
fraction of long-duration GRBs is directly connected with the death
of massive stars. Most evidence arises from observations of supernova
(SN) features in the spectra of a handful of GRB afterglows. Examples
of the spectroscopic SN/GRB connection include SN\,1998bw/GRB\,980425
(Galama et al. 1998), SN\,2003dh/GRB\,030329 (Stanek et al. 2003;
Hjorth et al. 2003), SN\,2003lw/GRB\,031203 (Malesani et al. 2004),
SN\,2002lt/GRB\,021211 (Della Valle et al. 2003), XRF\,020903 with a
broad-line type-Ib/c SN (Soderberg et al. 2005), and the recent
SN\,2006aj/GRB\,060218  (Masetti et al. 2006; Fatkthullin et
al. 2006; Campana et al. 2006; Modjaz et al. 2006).  In addition
there are about a dozen afterglows which show rebrightenings and/or
flattenings in their light curve days to weeks after the gamma-ray
event (e.g. Bloom et al. 1999; Zeh et al. 2004). These bumps are
interpreted as due to the emergence of SN. Most of these studies are
based upon photometric data, due to the extreme faintness of late-time
afterglows. There is some evidence that the population of SNe
associated with GRBs is quite heterogeneous: several bumps have light
curves which are well matched by faint SNe (e.g. GRB\,011121:
Garnavich et al. 2003, Bloom et al. 2002; XRF\,030723: Fynbo et
al. 2004; GRB\,020410: Levan et al. 2005; XRF\,040701: Soderberg et
al. 2005), rather than bright ``hypernovae'' such as SN\,1998bw. 
However, only spectroscopic observations can definitely assess the
degree of ``diversity'' among GRB progenitors.  Moreover, dust echos
(e.g. Esin \& Blandford 2000; Waxman \& Draine 2000) can produce 
red late-time `rebrightenings' resembling SN bumps, and spectroscopy
is the most effective way to identify their nature. Despite the
intense interest in detailing the diversity of GRB-SNe, only one
association with a \textit{Swift} burst has been reported to date
(GRB\,060218/SN\,2006aj). This is very likely due to the higher
redshifts of \textit{Swift} bursts ($\langle z \rangle = 2.8$;
Jakobsson et al. 2006) compared with pre-\textit{Swift} events. In
particular, only about 10\% of the almost 100 \textit{Swift} bursts
have occured at redshift less than one. At redshifts $\geq 1$, SNe are
almost impossible to detect with current instrumentation. Thus, the
discovery of low-redshift GRBs is a critical gateway for the study of
GRB-SNe. In this {\em Letter}, we detail the measurement of the
redshift of GRB\,050525A and present the discovery of an extra
component in its late afterglow, together with spectroscopic evidence
for an associated SN. This SN was dubbed SN\,2005nc \citep{IAUC}.

\section{GRB\,050525A}

The long-duration GRB\,050525A was discovered by the \textit{Swift}
satellite (Gehrels et al. 2004) on 2005 May 25.002 UT (Band et
al. 2005). It was a bright event, with fluence $\mathcal{F} = (2.01
\pm 0.05) \times 10^{-5}$~erg~cm$^{-2}$ and duration $T_{90} = 8.8 \pm
0.5$~s (this is the time during which 90\% of the photons are
collected). The burst was observed also by numerous other satellites
(INTEGRAL: G\"otz et al. 2005; \textit{Wind}: Golenetskii et al. 2005;
RHESSI; Mars Odyssey), allowing for a detailed study of the high energy
prompt emission spectrum. The optical afterglow was soon discovered by
the ROTSE-III robotic telescope (Rykoff et al. 2005a), as a bright,
fading source at $\alpha_{\rm J2000} = 18^{\rm h} 32^{\rm m} 32\fs6$,
$\delta_{\rm J2000} = +26\degr 20\arcmin 23\farcs5$ (Rykoff et
al. 2005b). The optical and X-ray counterparts were also monitored in
great detail by the X-Ray Telescope (XRT) and the Ultraviolet/Optical
Telescope (UVOT) onboard \textit{Swift} (Blustin et al. 2005). The
brightness of the afterglow allowed an intense monitoring of this
object, from optical (e.g. Klotz et al. 2005; Mirabal et al. 2005) to
radio frequencies (Cameron \& Frail 2005), including the first detection
of a GRB at mid-infrared wavelengths (3.6 to 24~$\mu$m; Garnavich et
al. 2005).

\begin{figure}\centering
\includegraphics[width=\columnwidth]{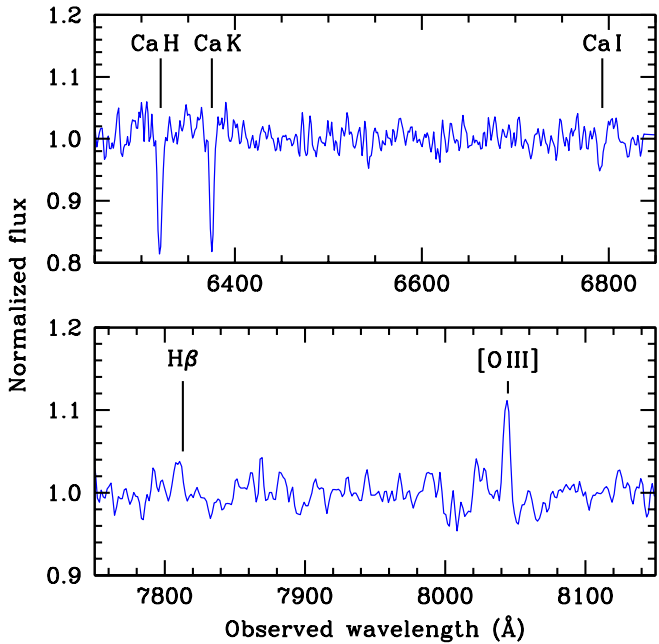}
\caption{Normalized spectrum of the afterglow of GRB\,050525A
on 2005 May 25.4~UT, obtained on Gemini North with the GMOS
instrument. \label{fg:spec}}
\end{figure}

\section{Observations and Data Analysis}

A spectrum of the afterglow was obtained with the Gemini North telescope
equipped with the GMOS instrument. The spectroscopic redshift was
measured through both emission and absorption lines, yielding $z =
0.606$ (Foley et al. 2005; see also Fig.~1). This is at the low end of
the GRB redshift distribution (Bloom et al. 2003, Berger et al. 2005,
Jakobsson et al. 2006) and is the third lowest redshift
\textit{Swift} long-duration GRB aside GRB\,050803 (Bloom et al. 2006)
and the recent GRB\,060218 (Mirabal \& Halpern 2006). This event
therefore presents a good opportunity for detailed analysis and
modelling. We observed the optical afterglow of GRB\,050525A with the
VLT and TNG telescopes, during the period 2005 May-September (from a few
hours up to $>100$~d after the gamma-ray event, see Fig.~2). Data
reduction was carried out following standard procedures. The photometric
calibration was achieved by observing several standard star fields on
different nights, yielding a zeropoint accuracy of $\approx
0.02$~mag. Photometric data (Fig.~2) show a flattening of the light
curve at $R \sim 24$, starting about 5~d after the burst (observer rest
frame) and lasting for about 20~d. The contribution of the host galaxy
during this phase is $\la 40\%$, as estimated from our
late-epoch images which show the host magnitude is fainter than $R \sim
25$. The afterglow contribution, as extrapolated from the earlier
measurement, is negligible at these epochs ($< 3\%$ at 20~d after the
GRB). This fact suggests that the flattening is powered by an additional
source of energy. To further quantify this, we fitted the observed light
curve including the contribution from the afterglow (a broken power
law), the host galaxy, and a SN component:
\begin{equation}
  F(t) = \frac{2F_{\rm b}}{[(t/t_{\rm b})^{\kappa\alpha_1}+(t/t_{\rm b})^{\kappa\alpha_2}]^{1/\kappa}}
       + F_{\rm h} + F_{\rm SN}[s(t-t_0)].
\end{equation}
Here $\alpha_1$, $\alpha_2$ (the early- and late-time slopes) and
$t_{\rm b}$ (the break time) characterize the afterglow decay ($\kappa$
determines how sharp is the break). $F_{\rm h}$ and $F_{\rm SN}$ are the
host galaxy and the SN flux, respectively. We allowed for a time offset
$t_0$ between the SN and the GRB, for a temporal stretch $s$ in the
light curve, and for a magnitude difference. For the SN flux, we adopted
the light curves of several type-Ic SNe (SN\,1998bw: Galama et al. 1998;
SN\,2002ap: Yoshii et al. 2003, Pandey et al. 2003, Gal-Yam et al. 2002,
Foley et al. 2003; SN\,1994I: Richmond et al. 1996). Our approach is
similar to that adopted by Zeh et al. (2004). The afterglow component is
well fitted with $\alpha_1 = 1.1$, $\alpha_2 = 1.8$, and $t_{\rm b} =
0.3$~d. Interpreting the break as due to a jet effect, this value of
$t_{\rm b}$ makes GRB\,050525A consistent with the Ghirlanda relation
(Ghirlanda et al. 2004; Nava et al. 2005). In Fig.~2, the solid line
shows the best fit using SN\,1998bw as a template, dimmed by
0.9~mag. The Galactic extinction towards GRB\,050525A is $A_R =
0.25$~mag (Schlegel et al. 1998). The extinction inside the GRB host is
uncertain. Blustin et al. (2005) suggest $E(B-V) \sim 0.1$ assuming an
SMC extinction curve, so that the observed $R$ band (roughly
corresponding to the rest-frame $B$) suffers an additional $A'_R \sim
0.35$~mag. This implies that the SN associated with GRB\,050525A may
have a luminosity just about 0.3~mag fainter than SN\,1998bw. However
fits of comparable quality (both in terms of visual appearance and
$\chi^2$) can be obtained by assuming other SN
templates. Figure~\ref{fg:lc_bump} shows fits obtained with the
hypernova SN\,2002ap (brightened by 1.3 mag) and the ``standard''
type-Ic SN\,1994I (brightened by 1.8~mag). Finally, we note that the
best matches to the data points are obtained by allowing a time offset
between a 1998bw-like SN and the GRB of $\approx 6$~d (corresponding to
$\approx 3.5$~d in the GRB rest frame) or a time stretch factor $s =
0.7$. These fits reproduce the flat shape of the rebrightening. A more
exotic scenario calls for the effect of a light echo. Indeed, sudden
flattenings have been observed in the light curves of SNe due to the
occurrence of light-echo components (e.g. Schmidt et al. 1994;
Cappellaro et al. 2001; Quinn et al. 2006, in preparation) during the
late SN decline.  However in these cases the flattenings were
characterized by much longer evolutionary time scales ($\sim $700~d)
than is observed for GRB afterglow bumps ($\sim 10$--20~d). In the SN
1991T echo-light, the spectrum at late stages was similar to that
emitted by the SN at maximum light, although about 10 mag fainter. In
our case, if the bump was due to an echo, we would expect to observe, in
the spectrum obtained during the flattening, the signatures (at
considerably fainter level of luminosity) of the very early optical
afterglow.  All of this indicates that a good set of photometric data
{\sl alone} may not constrain the SN type unambiguously. This task can
be tackled only by means of spectroscopic observations.

\begin{figure}\centering
\includegraphics[width=\columnwidth]{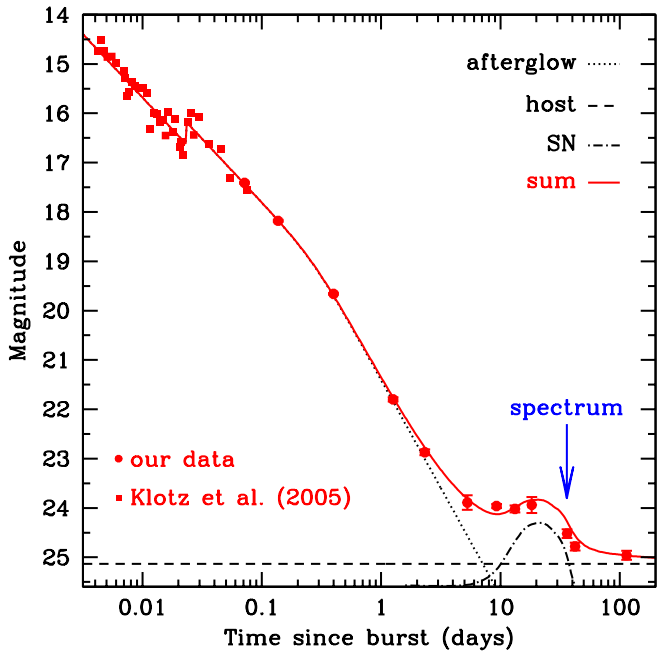}
\caption{$R$-band light curve of the afterglow of GRB\,050525A. Filled
circles represent our observations (TNG, VLT). Squares are from Klotz et
al. (2005), and were not included in the fit. The dotted, dashed and
dot-dashed lines indicate the afterglow, the host galaxy, and the SN
contributions, respectively.\label{fg:lc_all}}
\end{figure}

\begin{figure}\centering
\includegraphics[width=\columnwidth]{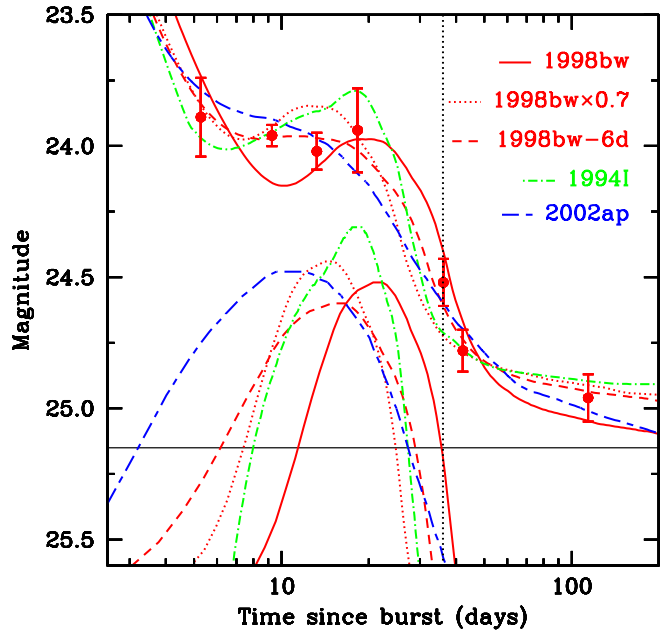}
\caption{Fits to the data points by using different SN templates.
 The solid horizontal line shows the contribution of the host galaxy.\label{fg:lc_bump}
The vertical dotted line represents the epoch of the spectrum.}
\end{figure}

\section{Spectroscopy}

\begin{figure}\centering
\includegraphics[angle=270,width=\columnwidth]{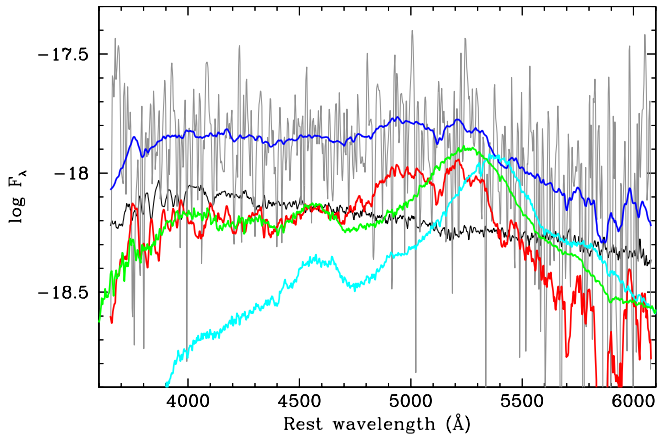}
\caption{Grey line: observed spectrum of GRB\,050525A, obtained during
the bump. Blue line: rebinned spectrum. Black line: template spectrum
for a blue starforming galaxy. Red line: subtracted spectrum
(blue$-$black). Green line: spectrum of SN\,1998bw 5d past
maximum. Cyan line: spectrum of SN\,1998bw 10d past
maximum.\label{fg:spec_SN}}
\end{figure}

A spectrum was obtained at the ESO VLT-UT1 with the FORS\,2 instrument
on 2005 Jun 28 (36~d after the burst, observer frame). The original
spectrum ($3 \times 40$~min integration) covered the range
5000--10000~\AA{} with a resolution of $\sim 20$~\AA. The extraction of
the spectrum was performed inside the IRAF and MIDAS environments. This
spectrum, shown in the rest-frame of the GRB as a grey line
(Fig.~\ref{fg:spec_SN}), was smoothed with a 100~\AA{} wide boxcar
filter (blue line) and cleaned from the host galaxy emission
features. As indicated by the light curve, this spectrum contains a
significant contribution from the host galaxy. We thus subtracted from
our spectrum a template of a blue star-forming galaxy (black line),
normalized to the host brightness and $R-I$ color. The use of a template
(average of 70 blue star-forming galaxies, from Cimatti et al. 2002),
was necessary due to the faintness of the host galaxy. An unsuccessful
attempt to get its spectrum with VLT was made on 2005 Oct 1. This
observation allowed us to set a robust upper limit to the magnitude of
the host ($R = 25.2 \pm 0.1$). The use of a blue star-forming galaxy
template is fully justified by several studies, which allow to conclude
that GRB hosts belong to this type of systems (e.g. Djorgovski et
al. 1998; Fruchter et al. 1999; Le Floc'h et al. 2003; Christensen et
al. 2004). The resulting spectrum (red line) is characterized by broad
undulations resembling the spectra of SN\,1998bw (Patat et al. 2001),
obtained $\sim 5$~d past maximum (green line). The best match is
obtained by dimming SN\,1998bw by $\Delta m \approx 0.9$ mag%
\footnote{The use of a template slightly brighter or fainter than
$R = 25.2$ shifts the date of the best match with SN\,1998bw by about
2~d later and earlier, respectively.}.  The bump at $\sim 5000$~\AA{}
has some contribution from residuals of sky emissions that we were
unable to remove. For comparison, we also show the spectrum of
SN\,1998bw $\sim 10$~d past maximum (cyan line). Inspection of
Fig.~\ref{fg:spec_SN} reveals a poor match with the observed data,
suggesting that SN\,2005nc was not far from maximum when the spectrum
was acquired. This result does not support the best fit to the
photometric data, which suggested either a time lag between the SN and
the GRB or a stretch factor $< 1$. Indeed, in these cases the spectrum
would have been obtained about 10~d (rest frame) past maximum, which is
excluded by spectroscopic observations (see Fig. 4). Spectra of SN 1994I
(Filippenko et al. 1995) obtained close to maximum light also do not
match the observed spectrum.

\section{Discussion}

Photometric and spectroscopic observations of the rebrightening
associated with GRB\,050525A support the idea that the bump was powered
by an emerging SN (SN\,2005nc). While photometric data alone are
consistent (within the errors) with different SN morphologies (based on
either SN\,1998bw, SN\,2002ap, or SN\,1994I templates), the analysis of
the spectrum obtained during the bump allowed us to select a SN akin to
SN\,1998bw as the best match, possibly dimmed by $\sim 0.3$~mag. The
flat shape of the bump light curve implies that SN\,2005nc brightened
relatively quickly after the GRB, with a rise time $\tau \sim 10$--12~d
in the $B$ band (for SN\,1998bw, $\tau_B \sim 14~d$). This can be
explained either by introducing a significant stretch factor ($<1$) or a
time lag between the SN and the GRB. Both possibilities would imply that
at the epoch of our spectrum (36~d after the GRB) the SN provided a very
small contribution to the observed light. This is inconsistent with our
observations, which suggest that the SN and the host galaxy had
comparable brightness ($R \sim 25.2$) at that epoch. We thus conclude
that SN\,2005nc had a different light curve from the templates we
adopted. Differences among GRB-SNe were noticed before the present study
(e.g. Cobb et al. 2004). We find that the SN associated with 050525A had
a faster rise than SN\,1998bw, but that it was characterized by a
broader, long-lasting maximum. On the other hand Maeda et al. (2005)
showed that asymmetric SNe peak earlier if observed close to their polar
axis. GRB\,050525A was a bright event, likely observed on-axis. On the
contrary, SN\,1998bw may have been observed off-axis (Yamazaki et
al. 2003), so that an earlier peak for SN\,2005nc could be expected in
this scenario.  In spite of the putative existence of a broad range in
the magnitudes at maximum of SNe associated with GRBs, all
GRB-associated SNe which have so far had spectroscopic confirmation (see
Della Valle 2006 and Woosley \& Bloom 2006 for reviews) appear to belong
to the bright tail of the type-Ib/c SN population (all objects mentioned
in the Introduction have $M_B \sim -19$). Whether this is the effect of
an observational bias (which favors the spectroscopic observations of
bright SNe) operating on a small number of objects, or whether it has a
deeper physical meaning is not yet clear. In this paper we have found a
new case of association between a GRB and a bright hypernova, both on
robust photometric grounds and spectroscopic evidence. On the other hand
to prove {\it beyond doubt} the existence of a class of GRB-associated
SNe having spectroscopic properties dramatically different from those
exhibited by the prototypical hypernova SN\,1998bw still remains one of
the most important observational challanges for the SN/GRB community,
that
\textit{Swift} will hopefully allow us to address in the near future.
\smallskip

M.D.V. thanks KITP at UCSB, where this work was completed, for its
hospitality. This research was supported in part by the National Science
Foundation under Grant No. PHY99-0974.  J.S.B., J.X.P., and H.-W.C. are
partially supported by NASA/\textit{Swift} grant NNG05GF55G. K.H. is
grateful for support under NASA grant FDNAG5-9210.  Based on
observations obtained at ESO, TNG and Gemini Observatories.  We wish to
extend special thanks to those of Hawaiian ancestry on whose sacred
mountain we are privileged to be guests.

%

\begin{table*}[!h]
\caption{Log of our $R$-band photometry.}
\centering
\begin{tabular}{llllll} \hline
Mean time (UT) & Time since GRB (d) & Telescope & Exposure time (s) & Seeing (\arcsec) & $R$ magnitude  \\ \hline
May 25.073     & 0.072              & TNG+LRS   & 2$\times$120      & 1.2              & 17.41$\pm$0.02 \\
May 25.139     & 0.137              & TNG+LRS   & 2$\times$120      & 1.2              & 18.18$\pm$0.02 \\
May 25.402     & 0.400              & VLT+FORS2 & 6$\times$120      & 1.1              & 19.66$\pm$0.02 \\
May 26.262     & 1.260              & VLT+FORS2 & 20$\times$90      & 0.6              & 21.80$\pm$0.04 \\
May 27.326     & 2.324              & VLT+FORS2 & 20$\times$90      & 0.4              & 22.87$\pm$0.06 \\
May 30.280     & 5.278              & VLT+FORS1 & 12$\times$180     & 1.0              & 23.89$\pm$0.15 \\
Jun 03.279     & 9.277              & VLT+FORS2 & 10$\times$180     & 0.4              & 23.96$\pm$0.04 \\
Jun 07.242     & 13.240             & VLT+FORS1 & 10$\times$180     & 0.9              & 24.02$\pm$0.07 \\
Jun 12.302     & 18.300             & VLT+FORS2 & 10$\times$180     & 0.6              & 23.94$\pm$0.16 \\
Jun 30.155     & 36.153             & VLT+FORS2 & 20$\times$90      & 0.4              & 24.52$\pm$0.09 \\
Jul 06.190     & 42.188             & VLT+FORS2 & 10$\times$180     & 0.4              & 24.78$\pm$0.08 \\
Sep 07.121     & 105.119            & VLT+FORS2 & 8$\times$240      & 0.5              & 24.96$\pm$0.09 \\ \hline
\end{tabular}
\end{table*}


\begin{thebibliography}{}

\bibitem[Band et al.(2005)]{Band05} Band, D., et al. 2005, GCN Circ. 3466
\bibitem[Berger \& Soderberg (2005)] {bersor} Berger, E., \& Soderberg, A. 2005, GCN Circ. 4384
\bibitem[Bloom et al.(1999)]{Bloom99} Bloom, J. S., et al. 1999, \nat, 401, 452
\bibitem[Bloom et al.(2002)]{Bloom02} Bloom, J. S., et al. 2002, \apj, 572, L45
\bibitem[Bloom et al.(2003)]{Bloom03} Bloom, J. S., Frail, D. A., \& Kulkarni, S. R. 2003, \apj, 594, 674
\bibitem[Bloom et al.(2006)]{Bloom06} Bloom, J.S., Perley, D., Foley,
R., Prochaska, J.X., Chen, H.-W., \& Starr, D. 2006, GCN Circ 3758
\bibitem[Blustin et al.(2005)]{Blustin05} Blustin, A. J., et al. 2006, \apj, 637, 901
\bibitem[Cameron \& Frail(2005)]{Cameron05} Cameron, P. B., \& Frail, D. A. 2005, GCN Circ. 3495
\bibitem[Campana et al.(2006)]{Campana06} Campana, S., et al. 2006, \nat, submitted (astro-ph/0603279)
\bibitem[Cappellaro et al.(2001)]{Capp01} Cappellaro, E., et al. 2001, \apj, 549, L215
\bibitem[Christensen et al.(2004)]{Christensen04} Christensen, L., Hjorth, J., \& Gorosabel, J. 2005, \aap, 425, 913
\bibitem[Cimatti et al.(2002)]{Cimatti02} Cimatti, A., et al. 2002, \aap, 381, L68
\bibitem[Cobb et al.(2004)]{Cobb04} Cobb, B. E., Bailyn, C. D., van Dokkum, P. G., Buxton, \& M. M., Bloom, J. S. 2004,
\apj, 608, L93
\bibitem[Della Valle et al.(2003)]{DellaValle03} Della Valle, M., et al. 2003, \aap, 406, L33
\bibitem[Della Valle(2005)]{DellaValle05} Della Valle, M. 2006, Il Nuovo
Cimento, 28, 563 (astro-ph/0504517)
\bibitem[Della Valle et al.(2006)]{IAUC} Della Valle, M., Malesani, D., Benetti, S., Chincarini, G., Stella, L., \& Tagliaferri, G. 2006, IAUC 8696
\bibitem[Djorgovski et al.(1998)]{Djorgovski98} Djorgovski, S. G., Kulkarni, S. R., Bloom, J. S., Goodrich, R., Frail, D. A., Piro, L., \& Palazzi, E. 1998, \apj, 508, L17
\bibitem[Esin \& Blandford(2000)]{EsinBlandford00} Esin, A. A., \& Blandford, R. 2000, \apj, 534, L151
\bibitem[Fatkthullin et al. (2006)]{Fatkthullin06} Fatkthullin, T.A., et al. 2006, GCN Circ 4809
\bibitem[Filippenko et al.(1995)]{Filippenko95} Filippenko, A., et al. 1995, \apj, 450, L11 
\bibitem[Foley et al.(2003)]{Foley03} Foley, R. J., et al. 2003, \pasp, 115, 1220
\bibitem[Foley et al.(2005)]{Foley05} Foley, R. J., Chen, H.-W., Bloom, J. S., \& Prochaska, J. X. 2005, GCN Circ. 3483
\bibitem[Fruchter et al.(1999)]{Fruchter99} Fruchter, A. S., et al. 1999, \apj, 519, L13
\bibitem[Fynbo et al.(2004)]{Fynbo04} Fynbo, J. P. U., et al. 2004, \apj, 609, 962
\bibitem[Galama et al.(1998)]{Galama98} Galama, T. J., et al. 1998, \nat, 395, 670
\bibitem[Gal-Yam et al.(2002)]{GalYam02} Gal-Yam, A., Ofek, E. O., \& Shemmer, O. 2002, MNRAS, 332, L73
\bibitem[Garnavich et al.(2003)]{Garnavich03} Garnavich, P. M., et al. 2003, \apj, 582, 924
\bibitem[Garnavich et al.(2005)]{Garna05} Garnavich, P. M., et al. 2005, GCN Circ. 3532
\bibitem[Gehrels et al.(2004)]{Gehrels04} Gehrels, N., et al. 2004, \apj, 611, 1005 
\bibitem[Ghirlanda et al.(2004)]{Ghirla04} Ghirlanda, G., Ghisellini,
G., \& Lazzati, D. 2004, \apj, 616, 331
\bibitem[Golenetskii et al.(2005)]{Gole05} Golenetskii, S., Aptekar, R., Mazets, S., Pal'shin, V., Frederiks, D., \& Cline, T. 2005, GCN Circ. 3474
\bibitem[G\"otz et al.(2005)]{Gotz05} G\"otz, D., Mereghetti S., Mowlavi, N., Shaw, S., Beck, M., Borkowski, J., \& Lund, N. 2005, GCN Circ. 3472
\bibitem[Hjorth et al.(2003)]{Hjorth03} Hjorth, J., et al. 2003, \nat, 423, 847
\bibitem[Jakobsson et al.(2006)]{Jakob05} Jakobsson, P., et al. 2006, \aap, 447, 897
\bibitem[Klotz et al.(2005)]{Klotz05} Klotz, A., Bo\"er, M., Atteia, J. L., Stratta, G., Behrend, R., Malacrino, F., \& Damerdji, Y. 2005, \aap, 439, L35
\bibitem[Le Floc'h et al.(2003)]{LeFloch03} Le Floc'h, E., et al. 2003, \aap, 400, 499
\bibitem[Levan et al.(2005)]{Levan05} Levan, A., et al. 2005, \apj, 624, 880
\bibitem[Maeda et al.(2005)]{Maeda05} Maeda, K., Mazzali, P. A., \& Nonoto, K. 2005, \apj, in press (astro-ph/0511389)
\bibitem[Malesani et al.(2004)]{Malesani04} Malesani, D., et al. 2004, \apj, 609, L5
\bibitem[Masetti et al. 2006]{masetti06} Masetti, N., Palazzi, E., Pian,
E., \& Patat, F. 2006, GCN Circ 4803
\bibitem[Mirabal et al.(2005)]{Mirabal05} Mirabal, N., Bonfield, D., \& Schawinski, K. 2005, GCN Circ. 3488
\bibitem[Mirabal et al.(2006)]{MiraHalp06} Mirabal, N., Halpern, J., An, D., Thorstensen, J. R., \& Terndruf, D. M. 2006, \apj L, submitted
\bibitem[Modjaz et al.(2006)]{Modjaz06} Modjaz, M. 2006, \apj L, submitted (astro-ph/0603377)
\bibitem[Nava et al.(2005)]{Nava05} Nava, L., Ghisellini, G., Ghirlanda,
G., Tavecchio, F., \& Firmani, C. 2005, \aap, in press (astro-ph/0511499)
\bibitem[Pandey et al.(2003)]{Pandey03} Pandey, S. B., Anupama, G. C., Sagar, R., Bhattacharya, D., Sahu, D. K., \& Pandey, J. C. 2003, MNRAS, 340, 375
\bibitem[Patat et al.(2001)]{Patat01} Patat, F., et al. 2001, \apj, 555, 900
\bibitem[Richmond et al.(1996)]{Richmond96} Richmond, M. W., et al. 1996,  \aj, 111, 327
\bibitem[Rykoff et al.(2005a)]{Rykoff05a} Rykoff, E., Yost, S. A., \& Swan, H. 2005a, GCN Circ. 3465
\bibitem[Rykoff et al.(2005b)]{Rykoff05b} Rykoff, E., Yost, S. A., Swan, H., \& Quimby, R. 2005b, GCN Circ. 3468
\bibitem[Schlegel et al.(1998)]{Schlegel98} Schlegel, D. J., Finkbeiner, D. P., \& Davis, M. S. 1998, \apj, 500, 525
\bibitem[Schmidt et al. (1994)]{Schmidt94} Schmidt, B. P., Kirshner, R. P., Leibundgut, B., Wells, L. A., Porter, A. C., 
Ruiz-Lapuente, P., Challis, P., \& Filippenko, A. V. 1994, \apj, 434, L19
\bibitem[Soderberg et al.(2005)]{Soderberg05} Soderberg, A. M., et al. 2005, \apj, 627, 877
\bibitem[Stanek et al.(2003)]{Stanek03} Stanek, K. Z., et al. 2003, \apj, 591, L17
\bibitem[Waxman \& Draine(2000)]{WaxmanDraine00} Waxman, E., \& Draine, B. T. 2000, \apj, 537, 796
\bibitem[Woosley \& Bloom(2006)]{Woosley06} Woosley, S., \& Bloom., J. S. 2006, \araa, submitted
\bibitem[Yamazaki et al.(2003)]{Yamazaki03} Yamazaki, R., Yonetoku, D., \& Nakamura, T. 2003, \apj, 594, L79
\bibitem[Yoshii et al.(2003)]{Yoshii03} Yoshii, Y., et al. 2003, \apj, 592, 467
\bibitem[Zeh et al.(2004)]{Zeh04} Zeh, A., Klose, S., \& Hartmann, D. H. 2004, \apj, 609, 952

\end{thebibliography}
\end{document}